\documentclass[aip,reprint]{revtex4-1}
\usepackage{graphicx}
\usepackage{subcaption}
\usepackage[T1]{fontenc}
\draft 

\begin{document}


\title{A phase-modulation interferometer for intense, ultrashort, near infrared laser pulses} 



\author{S. Ganeshamandiram}
\author{M. Niebuhr}
\author{F. Richter}
\author{U. Bangert}
\author{G. Sansone}
\author{F. Stienkemeier}
\author{L. Bruder}
\affiliation{Institute of Physics, University of Freiburg, Hermann-Herder-Str. 3, 79104 Freiburg, Germany}


\date{\today}

\begin{abstract}
The investigation of coherent phenomena in strong-field processes requires interferometric measurement schemes with high selectivity to disentangle the complex nonlinear response of the system. 
Interferometers combining acousto-optical phase modulation with lock-in detection feature excellent dynamic range and highly selective detection, thus providing a promising solution. 
However, acousto-optical modulators (AOMs) cause several issues when operated with intense, ultrashort, near infrared (NIR) laser pulses. 
The AOMs introduce temporal and angular dispersion, self-phase modulation and reduced acousto-optic efficiency at NIR wavelengths. 
Here, we present an acousto-optical phase modulation interferometer design that solves these issues. 
The presented solutions pave the way for the investigation of strong-field processes with phase-modulated interferometry and are also useful to improve the performance of phase-modulation interferometers in other applications. 

\end{abstract}


\maketitle 

\section{Introduction}
The interaction of intense laser fields with matter gives rise to peculiar strong-field phenomena\,\cite{brabec_intense_2000, wolter_strong-field_2015}. 
Examples are the AC-Stark effect and Rabi oscillations or multiphoton, tunnel and above barrier ionization, and connected to this, high harmonic generation\,\cite{rabi_space_1937, autler_stark_1955, brabec_intense_2000}. 
The theoretical description of these phenomena can be challenging, in particular in complex quantum systems. 
On the one hand, the intense fields can induce a multitude of superimposing nonlinear processes leading to a highly complex system response. 
On the other hand, non-perturbative theory is required to capture the strong-field induced dynamics. 

To experimentally disentangle the strong-field response of a system and to gain a deeper understanding of strong-field phenomena, interferometry seems a suitable approach. 
Interferometry increases the attainable spectro-temporal resolution and allows to coherently amplify subtle spectroscopic signatures.  
Moreover, the mutual relation between time- and frequency interferograms offers complementary information on the investigated process in a single measurement.

Various interferometer designs have been developed to provide the required high phase stability for interferometry in the visible wavelength domain\,\cite{scherer_fluorescencedetected_1991, cowan_two-dimensional_2004, brixner_phase-stabilized_2004, tian_femtosecond_2003, rehault_two-dimensional_2014, tekavec_wave_2006, vaughan_coherently_2007}. 
However, the investigation of strong-field processes poses additional challenges. 
One is, that the optical setups need to cope with intense laser fields comprising typically of multi-mJ, sub-100 fs optical pulses. 
The other is that the strong fields induce a complex nonlinear response that is difficult to disentangle. 
This requires interferometry featuring selective detection schemes and very high dynamic range. 

The interferometric phase modulation (PM) technique\,\cite{tekavec_wave_2006} shows promise to solve these issues. 
This technique combines rapid acousto-optical PM with lock-in detection, which provides both selective detection and exceptionally high dynamic range, paired with excellent phase stability. 
The advantages of the PM technique have been demonstrated in various time-domain interferometry experiments in the visible (VIS) and even up to the extreme ultraviolet (XUV) domain\,\cite{wituschek_tracking_2020, uhl_improved_2022, wituschek_phase_2020}. 
Furthermore, the method can be readily extended to nonlinear interferometry probing higher-order multi-particle or multiphoton quantum coherences\,\cite{bruder_efficient_2015, bruder_phase-modulated_2017, bruder_delocalized_2019, yu_observation_2019, autry_excitation_2020} or probing multidimensional frequency-correlations\,\cite{tekavec_fluorescence-detected_2007, bruder_coherent_2018, tiwari_spatially-resolved_2018, bangert_high-resolution_2022}. 
The method seems thus particularly suitable for the interrogation of coherent phenomena in strong-field processes and first theoretical proposals in this direction have been made\,\cite{granados_decoding_2024}. 

With a few exceptions\,\cite{flogel_rabi_2017, nandi_observation_2022, nandi_generation_2024, richter_strong-field_2024}, the majority of strong-field phenomena have been induced and studied with long-wavelength radiation\,\cite{brabec_intense_2000}, mainly with NIR fields for which the required intense laser sources are nowadays readily available. 
However, several technical challenges arise when a PM interferometer is operated with intense NIR laser pulses. 
The material of the acousto-optical modulators (AOMs) introduces substantial material dispersion and causes self-phase modulation (SPM), the diffracted laser beams suffer from angular dispersion, and the acousto-optical efficiency is rather low at NIR wavelengths. 
Several PM interferometers have been previously reported\,\cite{tekavec_wave_2006, widom_solution_2013, wituschek_stable_2019, javed_broadband_2024, jana_overcoming_2025} each targeted for specific applications. 
However, the mentioned challenges arising for operation with intense, ultrashort, NIR, multi-mJ laser pulses have not been solved in the previous designs. 
Here, we solve these issues with a custom interferometer design. 
The technical solutions presented in our work are not only interesting for potential application in strong-field experiments but may also be used to improve the general performance in PM interferometry for any other application.

\section{PM interferometry}
\begin{figure}
    \centering
    \includegraphics[]{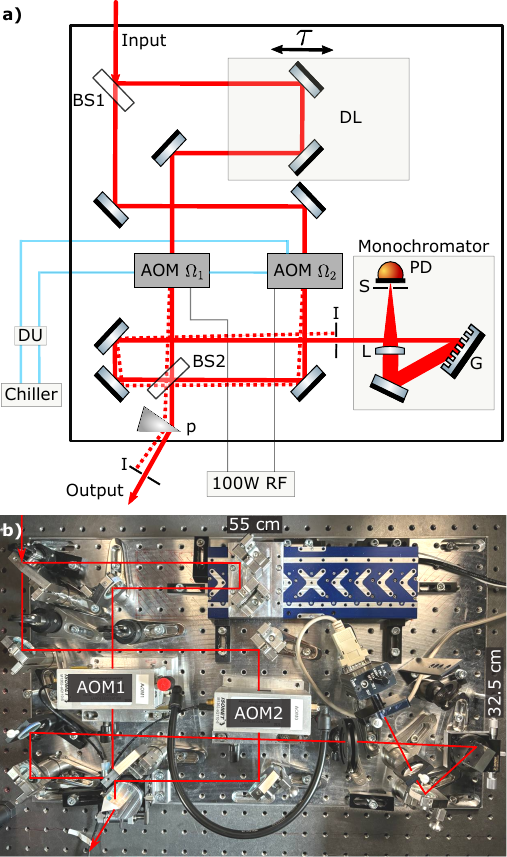}
    \caption{(a) Experimental setup consisting of a Mach-Zehnder interferometer equipped with a motorized delay line (DL) and two AOMs as well as a monochromator. The AOMs are driven at RF frequencies $\Omega_{1,2}$, respectively, and are water-cooled using a water chiller circuit. A damping unit (DU) is used to damp the vibrations in the water flow.
    Other abbreviations are: beam splitters (BS1/BS2), grating (G), lens (L), iris (I), photodiode (PD), slit (S), and prism (P). Red lines indicate the laser beam path and the dashed red lines the zero-order diffraction beams. (b) Photo of the experimental setup along with dimensions.}
    \label{fig:setup}
\end{figure}

First, we briefly describe the basic concept of the PM technique. 
For more details we refer to Ref.\,\cite{tekavec_wave_2006}. 
Fig.\,\ref{fig:setup}a shows the interferometer setup comprising of a Mach-Zehnder design with two AOMs placed in each interferometer arm. 
The AOMs shift the carrier-envelope phase (CEP) of each transmitted laser pulse. 
Phase-locked driving of the AOMs at distinct radio frequencies $\Omega_{1,2}$ thus induces a quasi-continuous low-frequency beat note at $\Omega_{21}=\Omega_2-\Omega_1$ between the two interferometer arms. 
Consequently, the interference signal $S_\mathrm{int}$ depends on the optical delay $\tau$ between both interferometer arms and, in addition, on the phase difference imprinted by the AOMs: 
\begin{equation}
    \label{1}
    S_\mathrm{int}(\tau,t) \propto \cos ( \omega_\mathrm{sig}\tau+\Omega_{21}t) \, ,
\end{equation}
where $\omega_\mathrm{sig}$ denotes the optical frequency and $t$ the laboratory time quasi-continuously sampled by the sampling rate of the experiment. 

This modulation approach has several advantages, that seem particularly beneficial for the investigation of the complex nonlinear response in strong-field experiments:

(i) The distinct modulation frequency $\Omega_{21}$ allows for efficient lock-in detection and amplification, which greatly enhances the dynamic range of the experiment. 
Ideally, $\Omega_{21}$ is chosen well above the 1/f noise spectrum of the laboratory environment (typ. $> 1$\,kHz) to ensure good suppression of noise and cross-talk with other signals. 
In case the experiment does not permit high signal sampling rates (e.g. due to low laser repetition rates or slow detectors), phase-synchronous undersampling can be implemented to maintain the SNR advantage\,\cite{bruder_phase-synchronous_2018}. 

(ii) The PM technique introduces a convenient way to efficiently extract and disentangle the nonlinear light-matter response. 
Nonlinear processes, e.g. multiphoton excitation, leads to a modulation at harmonics of the $\Omega_{21}$ frequency and can thus be filtered with high efficiency\,\cite{tian_ultrafast_2002, bruder_efficient_2015}. 
Similarly, multiple interferometers, each set to a different characteristic beat note, can be combined for coherent multidimensional spectroscopy\,\cite{tekavec_fluorescence-detected_2007,  bruder_coherent_2018, yu_observation_2019, autry_excitation_2020}. 
Here, the nonlinear signals are detected and separated from the linear background by amplifying specific sum- and difference-frequency combinations of the individual interferometer modulations. 

(iii) Phase noise picked up by the optical interferometer (e.g. due to pathlength fluctuations) is corrected upon the lock-in detection. 
To this end, an optical reference signal $R$ is coupled out at the second exit port of the recombining beamsplitter (BS2). 
This beam is spectrally filtered in a monochromator set to the optical frequency $\omega_\mathrm{ref}$, yielding
\begin{equation}
    \label{2}
    R(\tau,t) \propto \cos(\omega_\mathrm{ref}\tau+\Omega_{21}t).
\end{equation}
Phase-synchronous lock-in detection using this reference signal returns the demodulated signal $\bar{S}$ scaling with the phase difference between signal and reference: 
\begin{equation}
    \label{3}
    \bar{S}(\tau) \propto \cos([\omega_\mathrm{sig}-\omega_\mathrm{ref}]\tau) \, .
\end{equation}
This differential measurement leads to cancellation of the majority of phase noise, thus, providing a convenient passive interferometer stabilization. 

(iv) Rotating frame detection is introduced. $\bar{S}$ scales with the frequency difference $\omega_\mathrm{sig}-\omega_\mathrm{ref} \ll \omega_\mathrm{sig}$ and thus, the Nyquist limit is drastically lowered. 
This is especially of advantage in experiments involving short wavelengths, e.g. in the investigation of high-order harmonic generation processes, where otherwise very fine sampling with step sizes of $< 100$\,as is necessary\,\cite{wituschek_tracking_2020}. 

(v) The phase-synchronous detection enables the reconstruction of in-phase and in-quadrature interferograms from which the susceptibility of the studied system can be constructed. 
While in linear experiments the susceptibility may be readily calculated from the Kramers-Kronig relation, in highly nonlinear experiments the optical susceptibility can be more complex and direct experimental determination becomes important. 

These features make the PM technique particularly appealing for the investigation of strong-field phenomena and involved quantum coherences.

\section{Experimental Setup}
A schematic drawing of the interferometer setup along with a photo of the experimental setup is shown in Fig.\,\ref{fig:setup}. 
The Mach-Zehnder interferometer is equipped with two custom-manufactured AOMs (\textit{ISOMET M1362-aQ110-6}) and a magnetic-drive translation stage (\textit{Physik Instrumente, V-508.652020}). 
The AOMs feature a large aperture and low material dispersion to be compatible with intense, ultrashort laser pulses. 
The laser beam is s-polarized and the AOMs diffract in the horizontal dimension. 
Irises are used to block the zero-order diffraction of the AOMs and the $+1$-diffraction order is further propagated. 
The base plate of the AOMs is water-cooled to reduce the heat from sources inside the interferometer. 
A prism (details below) placed at the output of the interferometer compensates the angular dispersion introduced upon diffraction in the AOMs. 
We use a custom-built phase-locked dual-channel radio frequency (RF) driver to drive the AOMs at an acoustic frequency of 110\,MHz with a well-defined frequency difference $\Omega_{21}$ between the two AOMs. 
Depending on the experiment and the laboratory environment $\Omega_{21}$ is set to a value in the range of 10\,Hz to a few kHz. 
Further details about the AOM design and driver are stated in the results section. 

The interferometer has two exit ports at the recombining beamsplitter BS2. 
One is sent to the sample. 
Note, that for this exit port the material dispersion introduced by the beamsplitters is balanced between both interferometer arms. 
The second output of BS2 is sent into a home-built monochromator consisting of a reflection grating (G, 1800 grooves/mm), a plane mirror, a plano-convex lens (L, f=75 mm), a slit (S) and an amplified photodiode (PD).  
All other optics are designed for high damage threshold and low material dispersion: we use high-reflection dielectric mirrors (\textit{Layertec}, 100715) and beam splitters with 1\,mm-thick UVFS substrate (\textit{Layertec}, 141511). 
Eventually the optical aperture of the interferometer has a large diameter of 6\,mm to further reduce the laser intensity.

\section{Results and Discussion}
We characterized the performance of the interferometer setup with an amplified Ti:Sa laser system of which we expanded the beam to a diameter of 6\,mm. In tests where the spectral bandwidth is of particular importance we employed a non-collinear parametric amplifier (NOPA) delivering broad spectral bandwidth NIR pulses. Here we used the native beam diameter of $<$ 4\,mm. 
The relevant specifications of both systems are given for each measurement below. 

\subsection{Large aperture AOM design}

\begin{table*} 
    \centering
    \caption{Overview of different AOM materials and designs. 
    We compare our AOM design based on a quartz crystal with other common materials and crystal lengths used in PM interferometers, that are SF57\,\cite{tiwari_strongly_2018} and $\mathrm{TeO_2}$\,\cite{tekavec_wave_2006}. 
    Abbreviations: crystal length ($L$), nonlinear refractive index ($\mathrm{n_2}$), acoustic figure of merit ($\mathrm{M_2}$), group velocity dispersion (GVD), group delay dispersion (GDD). 
    $\mathrm{M_2}$ values are calculated using Eq.\,\ref{4a}. 
    The required photo-elastic coefficient $\mathrm{p}$ was only found for quartz\,\cite{ghatak_optical_1989} and $\mathrm{TeO_2}$\,\cite{souilhac_photoelastic_1989}. Both materials have very similar $\mathrm{p}$ values. Thus, as estimate for SF57 we assumed the same $\mathrm{p}$ value as for quartz. 
    Where relevant, the parameters are given for a wavelength of 800\,nm. 
    An exception is $\mathrm{n_2}$ of SF57, which was only found for a wavelength of 1064\,nm\,\cite{cimek_experimental_2017}.}
    \begin{tabular}{cccccc}
        \hline
        \hline
        Crystal material & $L$ (mm) & $\mathrm{n_2 \, (\times10^{-20} m^2/W)}$ & $\mathrm{M_2 \, (\times10^{-15} s^3/kg)}$ & GVD ($\, \mathrm{fs^2/mm}$) & GDD ($\, \mathrm{fs^2}$)\\ 
        \hline
         Quartz & 13 & $2.7 $ & $\mathrm{1.4}$ & 36 & 470\\
         SF57 & 11 & $13.9$ & $\mathrm{12.2}$ & 219 &  2410\\
         $\mathrm{TeO_2}$ & 6 & $58.6$ & $\mathrm{17.1}$ & 497 & 2982\\
         \hline
         \hline
    \end{tabular}
   
    \label{tab:materials}
\end{table*}

The design of the AOMs requires particular attention. 
The diffraction efficiency of an AOM is given by\cite{goutzoulis_principles_1994},
\begin{equation}
    \label{4}
    \eta \simeq \mathrm{\frac{\pi^2}{2\lambda_0^2\cos^2\theta_0} M_2 \frac{P_aL}{H}}
\end{equation}
where $\lambda_0$ is the optical wavelength, $\theta_0$ is the angle of the laser beam inside the crystal, $\mathrm{P_a}$ is the applied acoustic power, $\mathrm{H}$ is the height of the acoustic beam (the length of the acoustic wave traveling through the crystal), and $\mathrm{L}$ is the acousto-optic interaction length. 
Moreover, $\mathrm{M_2}$ is the acoustic figure of merit of the crystal, which is defined as\cite{goutzoulis_principles_1994},
\begin{equation}
    \label{4a}
    \mathrm{M_2} = \frac{\mathrm{n^6p^2}}{\mathrm{\rho V^3}} 
\end{equation}
where, n is the linear refractive index, p is the effective photo-elastic coefficient, $\rho$ is the density and V is the acoustic velocity in the crystal medium.

Due to the unfavorable scaling with $\lambda_0$, application in the NIR wavelength regime becomes difficult. 
This is usually counteracted with crystal materials that feature large acousto-optic efficiency (large $\mathrm{M_2}$), such as TeO$_2$ or heavy glasses (e.g. SF57). 
However, these materials introduce large material dispersion and can exhibit fairly low damage threshold. 
In addition, the laser beam is typically focused into the AOM crystal which allows using small active apertures (small $\mathrm{H}$) to optimize the efficiency. 
Hence, this approach is not applicable for operation with high-intensity laser sources. 
\begin{figure}
    \centering
    \includegraphics[]{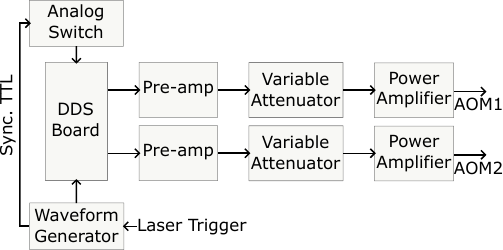}
    \caption{Schematic of custom-built RF driver used for phase-locked driving of the AOMs in pulsed opertaion mode.}
    \label{fig:RFdriver}
\end{figure}

In our setup we solve this problem. 
We employ AOMs with a large active aperture (12\,mm $\times$ 6\,mm) and apply unfocused laser beams with a beam diameter of 6\,mm. 
The AOM crystals are made of quartz, which features higher damage threshold, lower nonlinear index of refraction and much lower material dispersion than the other common AOM materials. 
We further minimize the material dispersion and self-phase modulation by using a fairly short crystal length of 13\,mm. 
Table\,\ref{tab:materials} compares the different AOM materials and designs used in PM interferometers. 
It clearly shows the advantage of our design in terms of nonlinearity and temporal dispersion. 
However, the chosen material also features a much lower $\mathrm{M_2}$ value. 
In combination with the large active aperture and the short crystal length this leads to a low overall acousto-optic efficiency. 

\begin{figure}
    \centering
    \includegraphics[scale=0.75]{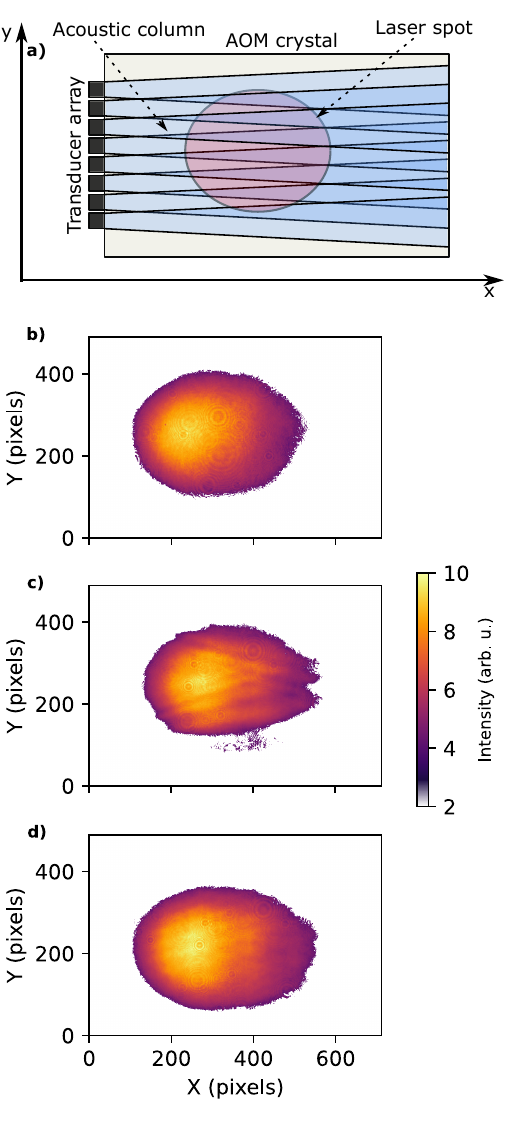}
    \caption{Beam profile distortion observed in the large aperture AOMs. (a) Schematic of the AOM cross section showing the laser beam transmitting the AOM crystal relative to the piezo-electric transducer position and the acoustic wave pattern in the crystal. 
    Comparison of beam profiles of the input beam entering the AOM (b), the first diffraction order of the AOM when transmitting the laser in the center of the AOM crystal (c) and after moving the laser beam as close as possible to the transducers (d).}
    \label{fig:beam_profiles}
\end{figure}

In principle, this can be counter-acted by applying a large acoustic power $\mathrm{P_a}$ to the AOMs. 
However, at a certain level this causes burning of the piezo transducers and introduces a detrimental heat source inside the interferometer. 
Our custom AOMs are rated for maximum 25\,W continuous acoustic power, which is higher than common AOMs but still insufficient to support high diffraction efficiencies. 
To solve this issue, we pulse $\mathrm{P_a}$ and synchronize the timing of the acoustic pulses with the arrival of the laser pulses in the AOMs, thereby keeping the average acoustic power low. 
To this end, we developed a custom radio frequency (RF) driver, consisting of a dual channel direct digital synthesizer (DDS) board (Analog Devices, AD9959) a preamplifier stage (Minicircuits, ZX60-33LN-S+), a variable attenuator (Minicircuits, ZX73-2500-S+) and a power amplifier (Minicircuits, ZHL-100W-52-S+) (Fig. \ref{fig:RFdriver}). 
The DDS board is clocked by a sine-wave generated by a waveform generator (RIGOL, DG992). 
For pulsed operation of the RF driver, the output amplitude of the DDS board can be controlled with an analog voltage supplied to dedicated analog input channels of the DDS board. 
In principle it would be sufficient to supply a TTL pulse sequence to these input channels in order to achieve pulsing of the RF driver output. 
However, in our setup we open and close the contacts of the input channels with a custom switch circuit (based on integrated circuit ISL84467), which enables a cleaner and faster pulsing of the RF signal. 
To this end, the second output channel of the waveform generator is synchronized to the laser trigger and is used to produce TTL pulses of adjustable duty cycle and lag time relative to the laser trigger. 
These TTL pulses are fed into the switching circuit to control the pulsing of the RF driver channels. 
The lag time is adjusted in a way that the acoustic pulse in the AOM crystal is synchronized with the laser pulses traveling through the AOM. 
Note that moving the laser position relative to the transducer position in the AOM (compare Fig.\,\ref{fig:beam_profiles}a) may require readjustment of the lag time.  

\begin{figure}
    \centering
    \includegraphics[]{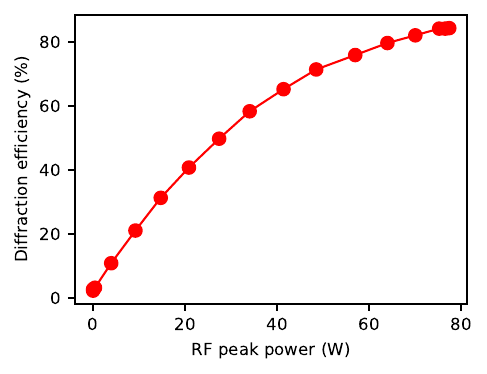}
    \caption{Diffraction efficiency of the AOM as a function of applied RF peak power when operated in pulsed-mode. Used RF pulse duration: 10\,\textmu s, duty cycle: 1\,\%.}
    \label{fig:efficiency}
\end{figure}
We tested pulsed operation for repetition rates of up to 200\,kHz, peak amplitudes up to 100\,W and duty cycles typically of a few percent. 
The residual average acoustic power leads still to moderate heating of the AOMs, which is dissipated by water cooling of the AOM base plate. 
To avoid any vibrations induced by the cooling water flow, we integrate a damping unit into the water supply circuit, as introduced in a previous setup\,\cite{wituschek_stable_2019}. 
Fig.\,\ref{fig:efficiency} shows the diffraction efficiency achieved in pulsed AOM operation for RF pulses of 10\,\textmu s duration and a duty cycle of 1\,\%. 
Laser pulses with center wavelength of 795\,nm and spectral bandwidth of 60\,nm (FWHM) were used. 
We achieve a diffraction efficiency of 84\,\% (ratio between diffracted laser power to input laser power) which is an excellent value at this wavelength given the special AOM design. 
The saturation of the curve implies that an optimum is reached at a RF peak power of 80\,W. 
For comparison, continuous-wave RF driving at the maximum rating of 25\,W would yield an efficiency of only $<$40 \,\%. 
The same diffraction efficiency is  also reached for laser pulses with larger spectral bandwidth (tested up to 80\,nm FWHM).

While the large active aperture AOM is beneficial to reduce the laser intensity inside the AOM crystals, the AOM design introduces also disadvantages. 
The large aperture of the acoustic column is created by an array of piezo-electric transducers rather than a single large area transducer (Fig.\,\ref{fig:beam_profiles}a). 
The acoustic wave generated by each of the transducers propagates through the crystal resulting in a spatial interference pattern in the laser beam cross section. 
This interference pattern imprints into the spatial intensity distribution of the diffracted laser beam, which we recorded with a CCD camera (compare input beam Fig.\,\ref{fig:beam_profiles}b to diffracted beam in c). 
Depending in which distance from the transducers the laser beam travels through the AOM, the interference pattern changes. 
We optimize the position to minimize the effect (Fig.\,\ref{fig:beam_profiles}d), which is for a laser beam position as close as possible to the transducers. 
In this way we obtain an interference contrast between both interferometer arms of $>$ 90\,\%, confirming that the wavefront of the diffracted beams are not seriously distorted by this AOM effect. 
For these measurements the Ti:Sa settings were: wavelength: 800\,nm, spectral width: 11\,\,nm (FWHM), pulse duration: 120\,fs. 

\subsection{Dispersion compensation}

\begin{figure}
    \centering
    \includegraphics[scale=0.8]{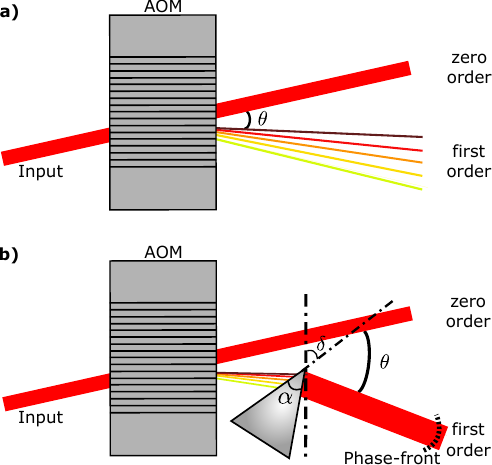}
    \caption{Schematic illustration of the angular dispersion produced by the AOMs: without using a compensating prism (a) and with a compensating prism (b).Diffraction angle: $\theta$, prism tilt angle: $\delta$, prism apex angle: $\alpha$, dashed curve: phase front. The phase front curvature caused by the prism is exaggerated for better visibility. }
    \label{fig:AOM_Prism_Scheme}
\end{figure}
In addition to the common material dispersion of the AOM crystal, the AOMs introduce an angular dispersion in the diffracted laser beams (Fig.\,\ref{fig:AOM_Prism_Scheme}a) according to
\begin{equation}
    \label{5}
    \frac{\theta}{\lambda} = \frac{\Omega}{v}
\end{equation}
where $\theta$ is the diffracted angle, $\lambda$ the optical wavelength, $\Omega$ the driving RF frequency and $v$ the acoustic velocity in the crystal (5.7 mm/\textmu s for quartz).
With this approximation we obtain an angular dispersion on the order of $\approx$19.3\,\textmu rad/nm. 
In applications where material dispersion is not of high concern, the AOMs can be operated in double-pass geometry, which doubles the material dispersion but compensates the introduced angular dispersion\,\cite{jana_overcoming_2025}. 
Alternatively, in low-intensity applications, this problem is partially compensated by focusing the laser into the AOMs, followed by collimation afterwards and propagation to the experiment. 
By focusing eventually the laser beam onto the spectroscopic target, this scheme maps in first order approximation the angular chirp introduced in the AOMs onto the interaction volume. 
Since the AOMs introduce a fairly small angular chirp, this procedure is in many applications tolerable. 
At least short pulse durations in the interaction volume are reported\,\cite{sahu_high-sensitivity_2023}, which confirms the small angular chirp in the laser focus.
\begin{figure*}
    \includegraphics[]{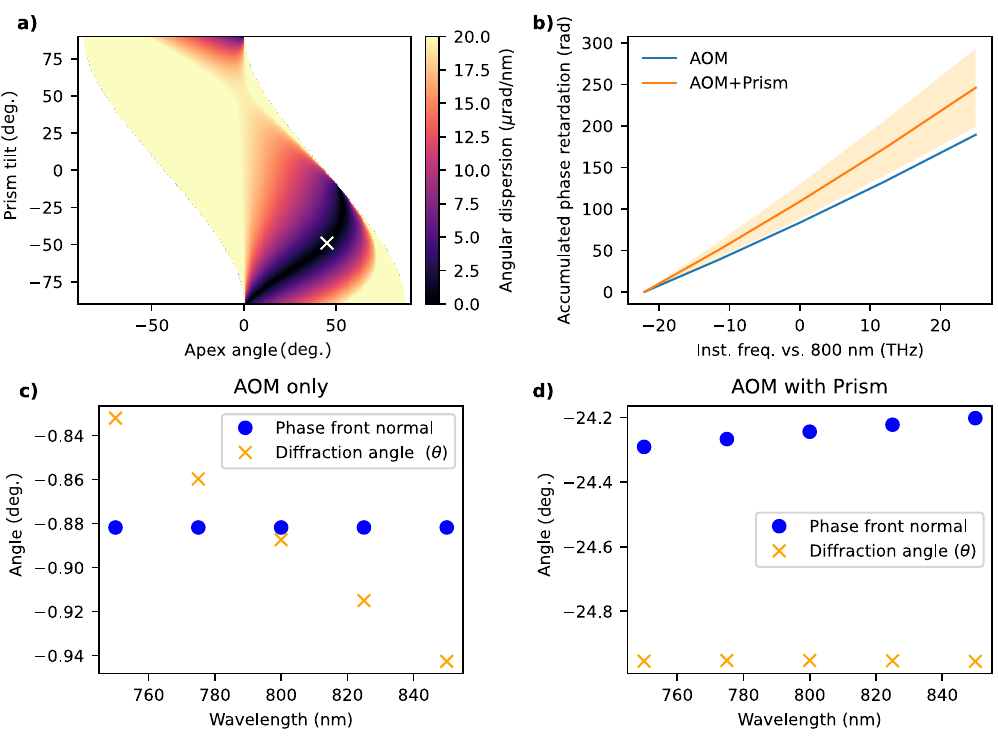}
    \caption{Ray tracing simulations of the diffraction in the AOM with and without including a compensation prism (see setup in Fig.\,\ref{fig:AOM_Prism_Scheme}). (a) Angular dispersion calculated for all possible prism tilt and apex angle combinations. Marked position (white cross) shows the chosen parameters in our setup. (b) Calculated accumulated phase in the beam passing through the setup in Fig.\,\ref{fig:AOM_Prism_Scheme} as a function of the instantaneous frequency given relative to the laser center wavelength of 800\,nm. For the setup without prism (blue) and with prism (orange). The orange shaded region shows the spread on the calculated phase retardation due to different lateral positions within the laser beam cross section. A polynomial fit yields the accumulated GDD at the center wavelength of 800\,nm. (c, d) Beam propagation angle $\theta$ (blue dots) and angle of the phase front normal (orange crosses) of the beam diffracted from the AOM  (compare Fig.\,\ref{fig:AOM_Prism_Scheme}). The two plots compare the case with and without inserting the compensation prism.}
    \label{fig:simulations}
\end{figure*}
Yet, these solutions are not suitable for high-intensity applications. 
Instead, in our setup we compensate the angular chirp with a prism placed in the diffracted output beam of the AOM (Fig.\,\ref{fig:AOM_Prism_Scheme}b, see also Fig.\,\ref{fig:setup}). 
The same approach is used in feed-forward CEP stabilization techniques\,\cite{lucking_long-term_2012}. 
To find the optimum prism material and apex angle we performed ray tracing simulations, from which we calculated the angular dispersion, the wave front tilt and the accumulated phase retardation in the diffracted laser beam (Fig.\,\ref{fig:simulations}). 
From the latter we derived the group delay dispersion (GDD) as a measure for the introduced temporal dispersion. 
To balance between the reflection losses, the diffracted beam diameter and the final propagation angle, we decided for an anti-reflection coated fused silica prism (Optik Krombach) with apex angle of 45$^\circ$ inserted into the setup at a tilt angle of -49$^\circ$ (white cross in Fig.\,\ref{fig:simulations}a).
Fig.\,\ref{fig:simulations}c,d show the comparison of the calculated angular dispersion and phase front tilt before/after inserting the prism into the ray tracing simulation. 
Clearly, without prism each wavelength propagates at a different angle, resulting in an angular dispersion of 19.2\,\textmu rad/nm. 
The value is very close to what is given by Eq.\,\ref{5}.
In this configuration, the phase front propagates perpendicular to the central wavelength at 800\,nm. 
Inserting the prism compensates the angular chirp very well but introduces a slight phase front tilt of 0.7$^\circ$ (Fig.\,\ref{fig:simulations}d). 
We note a residual variation of the phase front tilt of 0.09$^\circ$ across a bandwidth of 100\,nm, hence introducing a minor curvature of the phase front. 
Eventually, we computed the overall phase retardation introduced by the combination of the AOM and the prism as a function of optical frequency (Fig.\,\ref{fig:simulations}b). 
From this we derive the accumulated GDD to be 606\,fs$^2$ at 800\,nm at the center of the beam. However, we also estimated the accumulated GDD at the extreme positions on the beam to be 488\,fs$^2$ and 725\,fs$^2$. 
To check how much extra dispersion is introduced by the prism in our configuration, we compare our results with the dispersion caused by a quartz crystal with an equal length as the AOM crystal (13\,mm), which gives a value of GDD=496\,fs$^2$. 
This is a negligible difference in temporal dispersion at NIR laser wavelengths. 
Note, that for these calculations the prism was placed 2\,cm away from the AOM. 
Our calculations imply that the GDD accumulation decreases with the AOM-prism distance (not shown), hence increasing the AOM-prism distance will minimize the temporal dispersion of the laser pulses. However, the output beam waist increases with the AOM-prism distance (not shown).
We placed the prism closely behind the beamsplitter BS2 in our setup to balance between the temporal dispersion, the beam diameter, and the compactness of the interferometer. 

\begin{figure}
    \centering
    \includegraphics[scale=0.8]{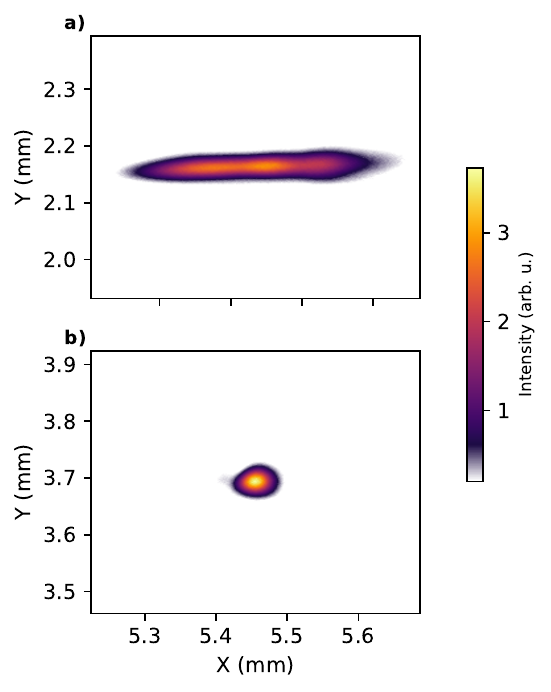}
    \caption{Comparison of the focal beam profiles when focusing the output beam of the interferometer with a lens ($f=200$\,mm) without using the compensation prism (a) and with the compensation prism (b).}
    \label{fig:focal_spot}
\end{figure}
As a first experimental validation of the angular chirp compensation by the prism, we compare how well the output beam of the interferometer can be focused. 
Without inserting the prism we obtain a strongly elongated focus due to the angular dispersion (Fig.\,\ref{fig:focal_spot}a), whereas with the prism, we get a symmetric focus (Fig.\,\ref{fig:focal_spot}b) of the same quality as for the input beam (not shown) entering the interferometer. 

\begin{figure}
    \centering
    
    \includegraphics[]{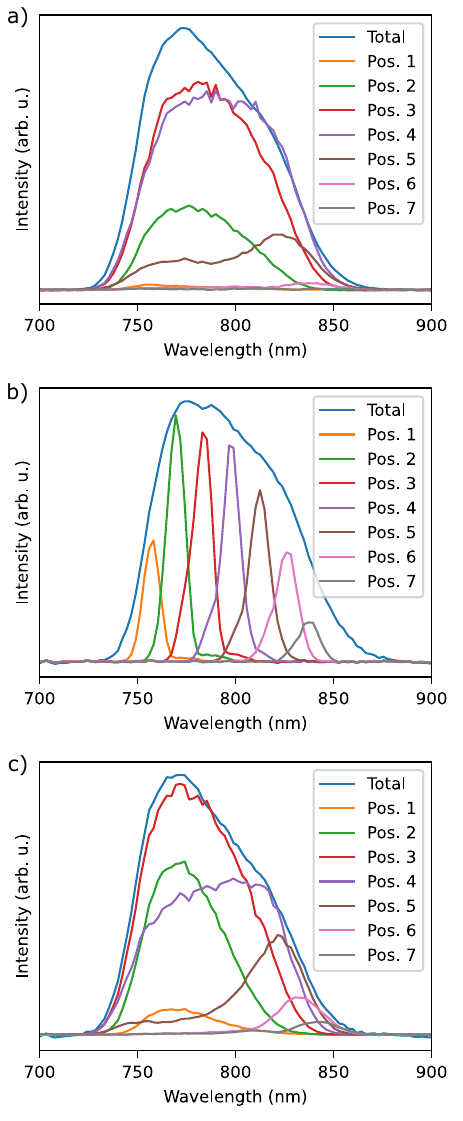}
    \caption{Investigation of the spatio-spectral characteristic in the laser focus. Spectra were recorded at different horizontal positions along the beam cross section by spatially select beam portions with a pin hole (25\,\textmu m diameter). (a) Input beam entering the interferometer, (b) output beam from the interferometer without using a compensation prism, (c) same but with using a compensation prism. For comparison the total spectrum is shown in each plot (blue trace) obtained by when not spatially select different beam positions.}
    \label{fig: ang.disp_scan}
\end{figure}
To analyze the angular chirp in more detail, we investigated next the spectro-spatial property of the laser focus (Fig.\,\ref{fig: ang.disp_scan}). 
To make this effect more visible, we used the spectrally broad NOPA output for these measurements (center wavelength: 790\,nm, bandwidth: $\approx$ 80\,nm (FWHM), pulse duration: 22\,fs, Fourier limit: 12\,fs, pulse energy: 0.9\,\textmu J, repetition rate: 50\,kHz). 
We focus the diffracted laser beam from the AOM and move a 25\textmu m pinhole along the focus cross section while recording the transmitted spectrum at each position. 
This gives an estimate of the spatial chirp in the laser focus which translates in first order to the angular chirp present in the unfocused beam. 
Fig.\,\ref{fig: ang.disp_scan}a shows the measurement for the input beam entering the interferometer, serving as a reference.  
We find a small spectral inhomogeneity across the laser focus, which is typical for the output of a NOPA when optimized for spectrally broadband pulses. 
After passing through the interferometer we find a strong position dependence of the spectral components in the focal cross section, which is reminiscent of the angular chirp introduced by the AOMs. 
Eventually, after inserting the prism this effect is well compensated and we obtain a residual spatial chirp comparable to the one present in the beam entering the interferometer. 
We show here only the horizontal axis which is the diffraction direction of the AOMs. 
In vertical direction we do not find a significant difference in the spectro-spatial beam property between input and output beam of the interferometer. 

\begin{figure}
    \centering
    \includegraphics[]{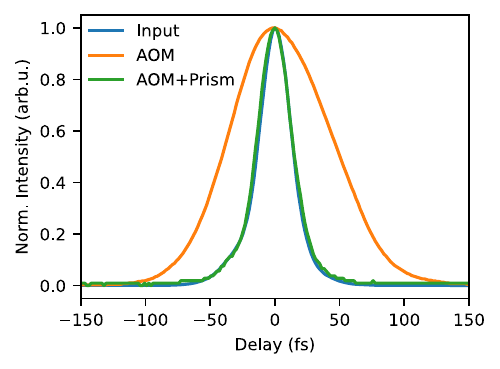}
    \caption{Second-harmonic autocorrelation measurements of the laser pulses entering the interferometer (blue) and of the output pulses from the interferometer when inserting the compensation prism (green) and without prism (orange).}
    \label{fig:FRAC}
\end{figure}
As a last validation, we used the same laser source and tested how well the output pulses from the interferometer can be temporally compressed. 
For compression we employ a prism compressor and for characterization we apply second-harmonic intensity autocorrelation measurements in a $\beta$-barium-borate crystal (crystal thickness: 20\,\textmu m). 
Fig.\,\ref{fig:FRAC} shows the result. 
Using the prism to compensate the angular dispersion, we are able to compress the output pulses of the interferometer to 22\,fs which is identical to the input pulses entering the interferometer. 
This is remarkable given the amount of material introduced in the interferometer. 
However, if we remove the prism, we manage to compress the output pulses only to 64\,fs, which is due to the angular chirp on the beam. 
We also compressed laser pulses from the amplified Ti:Sa laser featuring a Fourier-limit of 24\,fs and achieved compression to the same value after passing through the interferometer (not shown). 
This confirms, that by inserting the prism, we compensate spatial and temporal chirp effects sufficiently well to achieve short pulses. 

In overall, these tests show, that the issue of the angular and temporal dispersion introduced by the AOMs, can be adequately compensated with the insertion of a prism. 
While we chose a specific combination of apex and tilt angle for the prism, we find similar performance in our simulations for other prism dimensions (not shown).

\subsection{Performance of the interferometer}
\begin{figure}
    \centering
    \includegraphics[]{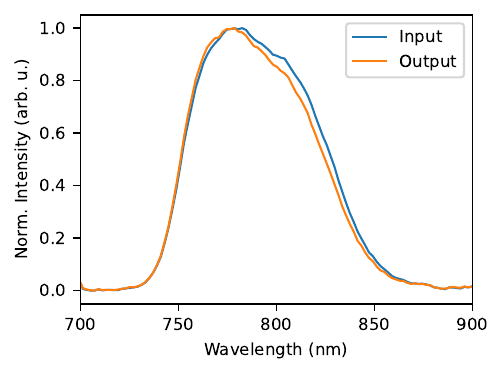}
    \caption{Spectral bandwidth acceptance of the interferometer. Input spectrum (blue), output spectrum (orange).}
    \label{fig:bandwidth}
\end{figure}
Eventually, we characterize the general performance of the PM interferometer. 
With the large-aperture AOM design discussed above we achieve very good diffraction efficiencies, which results in a good overall transmission of the interferometer. 
To characterize the optical losses, we measured the ratio between the input laser power and the combined output of both interferometer arms after the prism, from which we get a value of 42\,\% for the optical losses inside the interferometer. 
This is only slightly lower than the minimum loss of 50\,\% common to Mach-Zehnder and Michelson interferometers. 
In the previous section, we discussed the spatio-spectral beam property, however, without showing the overall spectral transmission of the interferometer setup. 
Fig.\,\ref{fig:bandwidth} compares the spectrum of broad bandwidth laser pulses at the input and output of the interferometer. 
Only a minor narrowing is observed in the output beam indicating the bandwidth limit imposed by the diffraction in the AOMs. 
The measurement confirms, that almost the full bandwidth of ultrashort laser pulses with Fourier-limited pulse duration of 12\,fs centered at around 790\,nm can be transmitted. 

Another important characteristic of the interferometer is the maximum acceptable laser intensity, which we determine from the onset of self-phase modulation observed in the laser pulses transmitted through the interferometer. 
Using the amplified Ti:Sa laser pulses, we find a discernible spectral broadening for an input peak intensity of roughly $>3.5\times10^{10}$\,W/cm$^2$. 
For even higher intensities we observe a reduction of the interference contrast pointing to a detrimental nonlinear effect in the AOM crystals. 
A detailed characterization of the latter phenomena will be subject of another publication. 
We furthermore investigated possibilities to operate the interferometer with pulses featuring even larger pulse energies. 
To this end, we stretched the input pulses temporally by a factor of 15 using the internal grating compressor of the Ti:Sa amplifier and post compress them after the interferometer with a chirped mirror compressor (5 mirror pairs, Ultrafast Innovations, HD31). 
We achieve good compression of the pulses to the same duration of 30\,fs of the input pulses while operating the interferometer with pulse energies of 1.7\,mJ without introducing discernible nonlinear effects. 
Even higher pulse energies should be possible in this setting, however, were not tested in our setup. 
\begin{figure}
    \centering
    \includegraphics[]{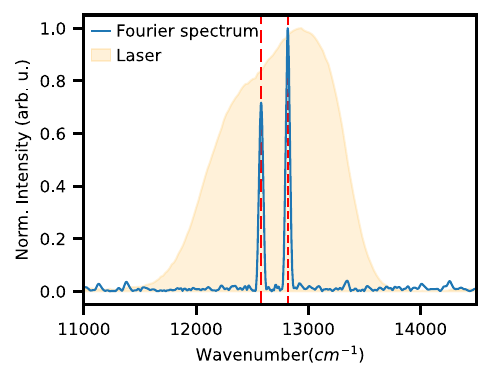}
    \caption{Fluorescence-detected Fourier-transform spectroscopy of a dilute Rb vapor recorded with the PM interferometer. Fourier spectrum (blue), laser spectrum (yellow shaded), positions of the D1 and D2 lines (red dashed), taken from NIST database\,\cite{kramida_a._nist_2015}.}
    \label{fig:Rb-Fl}
\end{figure}
As a final performance test of the PM interferometer, we performed a linear Fourier-transform spectroscopy experiment as described in Ref.\,\cite{tekavec_wave_2006}, using the NOPA as laser source. 
To this end, we focused (f=50\,mm) the output pulses into a spectroscopy cell, filled with a Rb atom vapor at room temperature. 
The laser-induced fluorescence is collected perpendicular to the laser beam propagation and detected with an amplified photodiode. 
The phase modulation of the interferometer arms leads to a modulation of the excited state populations in the Rb vapor, which is reflected in the fluorescence yield. 
The laser repetition rate is 50\,kHz and we chose $\Omega_{21}=300$\,Hz. 
The fluorescence signal is demodulated with a commercial lock-in amplifier using the reference signal detected in the home-built monochromator. 
For this purpose, the laser induced 50\,kHz-modulation of the reference signal is first removed with a low pass filter (cut off 5.3\,kHz). 
We scanned the relative pulse delay $\tau$ in the interferometer for a range of 50\,fs to 1.5\,ps with a step size of 5\,fs. 
At each delay position $\tau$ the signal was demodulated and averaged for 15000 laser shots. 
A discrete Fourier transform of the demodulated fluorescence with respect to $\tau$ yields the excitation spectrum of the sample, of which we show the absolute value in Fig.\,\ref{fig:Rb-Fl}. 
Two well-defined spectral peaks are observed, corresponding to the $5^2S_{1/2}\rightarrow5^2P_{1/2}$ (Rb D1 line) and $5^2S_{1/2}\rightarrow5^2P_{3/2}$ (Rb D2 line) transitions. 
The spectral distance of both peaks fits well with the literature. 
For the calibration of the absolute frequency axis, we compared the frequency of the D1 line with the value from the literature\,\cite{kramida_a._nist_2015} and calibrated the monochromator wavelength accordingly. 
The spectral resolution of 38.5\,cm$^{-1}$ is determined by the resolution of the home-built monochromator. 
This resolution should be sufficient for the investigation of most strong-field phenomena. 
In case, higher resolution is desired, the home-built monochromator may be replaced by a high-resolution monochromator readily available commercially. 
In overall, the Fourier-transform spectroscopy of a highly dilute vapor shows the general applicability of the interferometer setup.

\section{Summary}
The specific advantages of the PM technique over classical interferometry offers an interesting perspective to investigate strong-field phenomena and in particular their coherence properties. 
To this end, we have developed and characterized a PM interferometer design dedicated for the investigation of strong-field phenomena driven by ultrashort, multi-mJ pulses at NIR wavelengths centered around 800\,nm. 
Several challenges including the damage threshold and the spectral bandwidth acceptance of the interferometer as well as the temporal and angular dispersion introduced by the AOMs were addressed. 
The feasibility to perform interferometric measurements was demonstrated for the Fourier-transform type spectroscopy of a dilute Rb vapor. 

Our work opens up the investigation of strong-field processes such as above-threshold ionization and high-harmonic generation in complex systems for which the theoretical description is particular challenging and thus highly selective experiments based on the PM technique may be of great benefit. 
Moreover, the demonstrated design principle using short fused silica AOM crystals and pulsed RF operation will be beneficial to improve the performance of other PM interferometers employed also in other fields of research.


%
%

%

\begin{acknowledgments}
The authors acknowledge the funding from Deutsche Forschungsgemeinschaft RTG 2717, STI 125/24-1 and DFG SA3470/3-1, and European Research Council Starting Grant MULTIPLEX (101078689).
\end{acknowledgments}

\section*{Author Declarations}
\subsection*{Conflict of Interest}
The authors declare no conflict of interest.

\section*{Data Availability}
Data supporting the findings of this study are available from the corresponding author upon reasonable request.
\bibliography{Paper_Interferometer, 2025_NIRPM_inteferometer}

\end{document}